# Attosecond signatures in photodissociation by an intense Ti:Sapphire pulse

**J F McCann, L-Y Peng, I D Williams**

*International Research Centre for Experimental Physics, School of Mathematics and Physics, Queen's University Belfast, Belfast BT7 1NN, UK*

**Main contact email address:** *j.f.mccann@qub.ac.uk*

### Introduction

The creation of ultraintense coherent pulses of infrared light based on Ti:Sapphire lasers has produced trail-blazing explorations of the behaviour of materials exposed to extreme conditions. Energy deposition and transfer in atoms, molecules, clusters and solids, exposed to ultrashort pulses is of great current interest. Precise studies of energy transfer in molecules, leading to processes such as multiple photoionization, multiphoton dissociation and high-order harmonic generation have been successfully completed for a wide range of species[1]. The Central Laser Facility, in partnership with UCL, Strathclyde, Glasgow, Imperial College, Reading, St Andrews, Queen's University Belfast and other laboratories, has played a world-leading role in these studies within the Astra project.

One of the most important spectroscopic techniques used to analyse these interactions, is the measurement of the energy and angular distribution of fragment ions and electrons. For low-intensity, high-frequency light, it is well known that photoelectron spectroscopy is highly sensitive to molecular structure. One might suppose that high-intensity long-wavelength light would destroy this sensitivity. However, even under such extreme conditions, one can observe molecular 'fingerprints' in the fragments. Even more interestingly, these spectra can be a remarkably accurate means of calibrating the pulse itself. The aim of this theory paper is two-fold. Firstly, we make a direct comparison of our simulations with experimental work carried out recently at CLF[2] on proton emission spectra for intense-laser molecule photodissociation. Secondly, we examine the proton spectrum to determine whether it is possible to discern the signature of subcycle (attosecond) variations in an infrared pulse.

Although there exists a wealth of data on the photofragment energy spectrum for the neutral molecule, $H_2$, and its deuterated forms[1], the data for the much simpler isolated molecular ion is extremely scarce; hampered by the difficulty in preparing the target. In fact, experiments on $H_2^+$ in intense laser fields have only become possible in the last few years[2,3] through very recent advances in molecular beam technology.

### Description of the model

In a typical experiment[2,3] the $H_2$ molecules, at a temperature of a few hundred Kelvin, and hence predominantly in the vibrational state (v=0), can be converted into the bound molecular ion by electron-impact ionization,

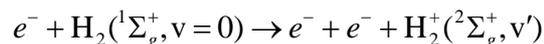

$$e^- + H_2(^1\Sigma_g^+, v=0) \rightarrow e^- + e^- + H_2^+(^2\Sigma_g^+, v')$$

The molecular ions can be extracted and collimated into a beam and injected at the focus of the laser. The manifold of vibrational v'-states is then subjected to intense Ti:Sapphire light, with wavelength λ ~750-800nm, and photon energy hν ~ 1.5eV. The photodissociation and/or photoionization processes,

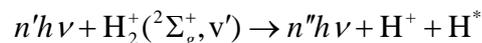

$$n'h\nu + H_2^+(^2\Sigma_g^+, v') \rightarrow n''h\nu + H^+ + H^*$$

produce ions and electrons that can be collected and analysed.

Owing to the small proton mass, the molecular expansion (vibration) is very rapid – on the scale of 5 fs or less. Since the rotational period of the molecular ion is rather long in comparison, 170 fs, a pulse lasting 60 fs or less, acts suddenly on the rotational timescale. One can assume the molecular axis as fixed, although randomly oriented, during this process. In our model, we establish the single molecule energy spectra, taking into account the pulse length and profile, and the orientation of the molecular axis, and average this data with the appropriate focal volume intensity variation in the experiment.

In our model the electron dynamics are described by a two-state approximation. This might appear a rough approximation since it neglects coupling through excited electronic states and ionization channels. However at high intensity (laser fields equivalent to the Coulomb force) and long wavelength (photon energy much lower than the ionization potential) the dissociation process occurs by adiabatic polarization involving the lowest electronic states. For the experiment in question, with intensities above $10^{14}$ W cm$^{-2}$ and photon energies of 1.5eV, such an approach is well justified. More contentious is the neglect of rotational heating (that is molecular realignment) during the pulse. However, this process is essentially an internal relaxation and for a 50fs pulse it has a small effect. To calculate the dissociation spectrum, we have applied discretization methods developed for photoionization of molecules[4] to solve the quantum equations in a dual configuration and momentum space. Technical details are discussed in the paper by Peng *et al.* [5].

### Proton emission spectroscopy

A very precise measurement of the angle and energy resolved dissociation spectrum was performed by Williams *et al.*[2] involving a team from Belfast, London and RAL. The effects of a linearly-polarized intense Ti:Sapphire (790nm) pulse interacting with a fast molecular ion beam were observed. Our simulations for this data, including focal volume sampling using the experimental parameters, is shown in Figure 1. Evidently, experiment and model are in remarkably good agreement.

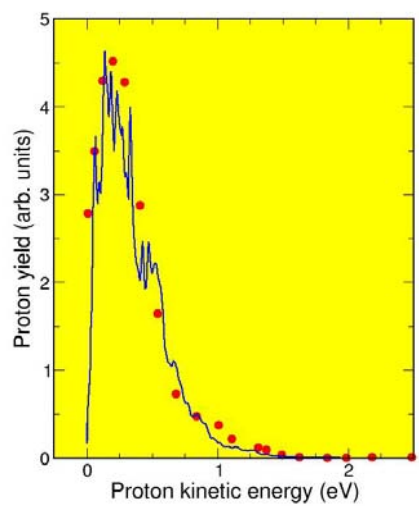

**Figure 1.** *Dissociation spectrum – theory vs expt.* Pulse intensity $I = 3 \times 10^{15}$ W cm$^{-2}$ with λ=790 nm and pulse length 65 fs. The solid blue curve is our calculation, the red circles the experimental measurements of Williams *et al.* [2] at the Central Laser Facility.

A striking feature of Figure 1 is that, in spite of the focal spot averaging process, the theoretical curve in Figure 1 exhibits sharp structures. This is one indication of the sensitivity of the proton yield to the laser intensity. Different regions of the focal spot contribute sharply defined proton spectra that survive averaging. Given these sharp variations, the excellent agreement of theory and experiment is more than fortuitous, since it relies on correct spatial and temporal averaging. A refinement of the model to include rotational heating would certainly smooth the sharpest theoretical peaks, but would have little effect on overall shape of the proton energy distribution. We conclude that our simple dynamical model is essentially correct and provides a very useful pulse calibration tool.

**Attosecond signatures**

The idea of using proton spectra to measure variations in the envelope of a 100fs pulse stems from the pioneering work of Codling and coworkers at Reading and the CLF at RAL[6]. Since dissociation is sensitive to the laser intensity and hydrogen stretch vibrations oscillate over a ~5fs timescale, it is clear that pulse duration and envelope variations over 10fs timescales can be resolved using proton spectroscopy (Figure 2). However, given that a Ti:Sapphire pulse cycle is ~3fs, which is short compared with the vibration time, it might seem unlikely that subcycle features of infrared light could be imprinted on the proton spectrum. However, in our limited study, we have found some evidence that optical pulse features at the subcycle (attosecond) timescale can be transferred through the electron-proton coupling to leave a signature in the proton spectrum (Figure 3).

Firstly, we consider the signatures of ultrashort pulse duration in the proton spectrum. The three vertical lines in Figure 2 indicate the expected energies of protons, in the centre-of-mass frame, arising from long, weak pulses of light. In Figure 2(a) we consider the proton spectrum from a 5fs pulse. Recall that a single cycle of 790nm light lasts only 2.6fs, so that such a short pulse has significant bandwidth (around 0.35eV in this case). The spectrum shows the presence of fast protons, with the energy distribution strongly bandwidth-broadened. A longer pulse, for example 20fs, allows the molecule to expand and then dissociate adiabatically giving slow protons, see Figure 2 (c). The spectra are quite distinct and, as expected, the proton spectrum provides a very good diagnostic of ultrashort pulse duration, and variations in the pulse envelope.

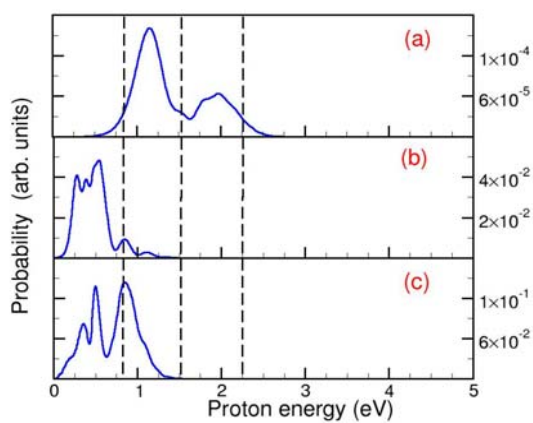

**Figure 2.** *Proton spectrum dependence on pulse length.* The figures show the transition from diabatic to adiabatic dissociation as the pulse length increases. In all cases the peak intensity was $I=5\times10^{14}$ W cm$^{-2}$ with $\lambda=790$ nm. The pulse duration (FWHM) is (a) 5fs (b) 10 fs and (c) 20 fs.

The molecular ion is created by vertical transition from the v=0 state of the neutral molecule, and consequently is a coherent mixture of the vibrational eigenstates. In Figure 3 (upper) we show the potential curve (dotted black line) of the molecular ion and the initial Franck-Condon wave packet (solid black line). Consider the field-free dispersion of this wave packet as shown by snapshots of the probability density at the beginning (solid black), and after 1fs (blue), 7fs (red) and 15fs (green). Clearly a 5fs pulse of light is short on this timescale so that a time delay (phase difference) between the molecular vibration and pulse arrival will be important. In our calculations of proton spectrum as a function of pulse delay this variation is very distinct, Figure 3 (lower). The coloured curves correspond to proton spectrum arising from the pulse delayed by 0fs (black) 1fs (blue), 7fs (red) and 15fs (green).

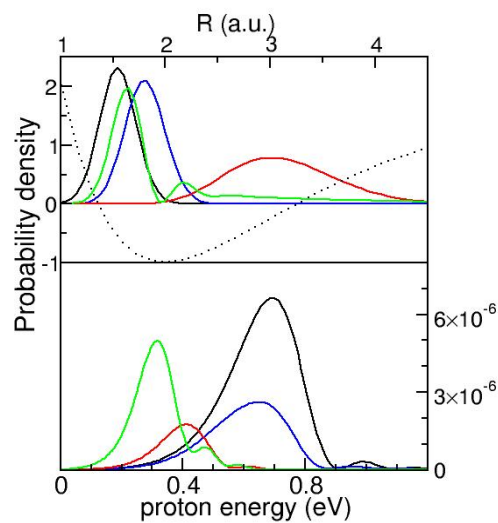

**Figure 3.** *Proton spectrum dependence on pulse delay.* The upper figure shows the unperturbed wave packet density as a function of time. The lower figure presents the proton emission spectrum resulting from pulse delay times: 0fs (black), 1fs (blue), 7fs (red) and 15fs (green). The proton spectrum resolves pulse delays at subcycle (attosecond) times.

The shortest delay times lead to fast protons, while a longer delay allows the molecule to stretch to where the dipole moment is strongest creating adiabatic dissociation with low energy protons. Even a very short delay (blue curve) gives a significant effect and thus allows some resolution at the subcycle limit. For an experiment, the observation of a time (phase) delay requires a mechanism to initiate (and calibrate) the molecular motion. Using a single laser pulse to photoionize and then dissociate, has the merits of controlling the sequence of these events. Recent advances in pulse compression have hinted at this possibility. It is hoped that accurate proton spectroscopy might soon yield evidence of subcycle pulse signatures in the proton spectrum from hydrogen groups in molecules.

**References**

1. J H Posthumus Rep. Prog. Phys., 67 623, (2004)

2. I D Williams, *et al.,* J. Phys. B 33 2743, (2000).
   K Sandig, *et al.,* Phys. Rev. Lett. 85 4876. (2000)

3. W R Newell, *et al.,* Euro. Phys. J. D 26 99 (2003)
   D Pavicic Euro. Phys. J. D 26 39 (2003)

4. L-Y Peng *et al.,* J. Phys. B 36 L295 (2003)

5. L-Y Peng *et al.,* J. Phys. B, 38 , 1727 (2005)

6. L J Frasinski *et al.,* Phys. Rev. Lett. 83 3625. (1999)